\documentclass[12pt]{article}
\usepackage{amsmath,amssymb,bm,epsfig,color}
\usepackage{psfrag}
\usepackage{epstopdf}

\renewcommand{\thefootnote}{\fnsymbol{footnote}}

\input epsf

\begin{document}

\title{
\begin{flushright}
\begin{minipage}{0.2\linewidth}
\normalsize
EPHOU-15-015 \\
WU-HEP-15-21 
\\*[50pt]
\end{minipage}
\end{flushright}
{\Large \bf 
New potentials for string axion inflation 
\\*[20pt]}}

\author{Tatsuo~Kobayashi$^{1,}$\footnote{
E-mail address:  kobayashi@particle.sci.hokudai.ac.jp}, \ \ 
Akane~Oikawa$^{2,}$\footnote{
E-mail address: a.oikawa@aoni.waseda.jp
} \ and \,
Hajime~Otsuka$^{2,}$\footnote{
E-mail address: h.otsuka@aoni.waseda.jp
}, \\*[20pt]
$^1${\it \normalsize 
Department of Physics, Hokkaido University, Sapporo 060-0810, Japan} \\
$^2${\it \normalsize 
Department of Physics, Waseda University, 
Tokyo 169-8555, Japan} 
\\*[50pt]}

\date{
\centerline{\small \bf Abstract}
\begin{minipage}{0.9\linewidth}
\medskip 
\medskip 
\small
We propose a new type of axion inflation 
with complex structure moduli in the framework of type IIB superstring theory compactified on the 
Calabi-Yau manifold. 
The inflaton is identified as the axion for the complex structure moduli whose potential 
is originating from  instantonic corrections appearing through the period vector of 
the mirror Calabi-Yau manifold. 
The axionic shift symmetry is broken down to the discrete one 
by the inclusion of the instantonic correction and certain three-form fluxes. 
Our proposed inflation scenario is compatible with  K\"ahler moduli stabilization. 
We also study a typical reheating temperature in the case of complex structure moduli inflation. 
\end{minipage}
}

\begin{titlepage}
\maketitle
\thispagestyle{empty}
\clearpage
\tableofcontents
\thispagestyle{empty}
\end{titlepage}

\renewcommand{\thefootnote}{\arabic{footnote}}
\setcounter{footnote}{0}
\vspace{35pt}

\section{Introduction}
The cosmic inflation is an attractive scenario which is tied to not only a thermal history 
of the Universe, but also the particle phenomenology. 
It is quite important to discuss the particle phenomenology and cosmology on the same footing. 
In particular,  superstring theory is  a good candidate for the unified theory of gravitational and 
gauge interactions, and gives us a tool to calculate the higher-dimensional operators which are sensitive to the 
fundamental scale.

In  higher-dimensional theory as well as superstring theory, the moduli fields associated with the metric, vector and tensor fields appear in its low-energy effective field theory. 
Since the scalar potential for these moduli fields are prohibited by the Lorentz and gauge symmetries, 
they are usually considered as candidates of inflaton fields. 
Moreover, in the type IIB superstring theory on the Calabi-Yau (CY) manifold, 
the author of Ref.~\cite{Giddings:2001yu} pointed out that 
the complex structure moduli and/or dilaton fields are generically stabilized 
by the flux-induced potential at the tree level,  
whereas the K\"ahler moduli can be stabilized at their minimum by certain nonperturbative 
effects. Thus, one of the K\"ahler moduli can play a role of the inflaton field.(See for review, e.g., Ref.~\cite{Baumann}.) 
The imaginary part of the K\"ahler moduli and scalar fields associated with the higher-dimensional  fields, 
i.e., the axions are also considered as candidates of the inflaton fields whose scalar potential is the form of natural inflation~\cite{Freese:1990rb} 
and/or axion-monodromy inflation~\cite{McAllister:2008hb,Silverstein:2008sg} . 
However, the complex structure moduli are not so discussed in the light of cosmic inflation 
and the thermal history of the Universe after the inflation. 
From the inflationary point of view, there are several studies in terms of complex structure moduli, for example,  
the natural inflation is derived from the nonperturbative effects through 
the threshold corrections~\cite{Abe:2014xja}\footnote{Through threshold corrections, 
one can realize a large decay constant enhanced by the inverse of loop factor. See also, Ref.~\cite{Abe:2014pwa}.}
 and instanton effects~\cite{Hebecker:2015rya}. 
The backreaction from the K\"ahler moduli is also discussed in Refs.~\cite{Hayashi:2014aua,Hebecker:2014kva}. 
As one of the difficulties to consider one of the complex structure moduli as the inflaton, one can 
expect that the K\"ahler moduli could be destabilized during and after the inflation 
because there appears a runaway direction in their potential.

In order to overcome them, we begin with the four-dimensional ($4$D) ${\cal N}=1$ supergravity 
derived from the type IIB string theory on the CY manifold. 
These instantonic effects for the prepotential of the CY manifold 
have been explicitly calculated by employing the technique of mirror symmetry 
within the framework of topological string~\cite{Witten:1988xj,Witten:1991zz}.
Therefore, the flux-induced superpotential, i.e., Gukov-Vafa-Witten (GVW) superpotential~\cite{Gukov:1999ya}, 
receives the worldsheet instanton effects. 
If the relevant complex structure moduli only appear from the nonperturbative effects through 
the period vector of the CY manifold, their energy scale can be smaller than that of other complex 
structure moduli and dilaton as well as the K\"ahler moduli. 
Then, these light complex structure moduli are stabilized at their 
minimum after the K\"ahler moduli stabilization. 
Thus, one can discuss the scenario where 
the inflaton is identified with  one of the complex structure moduli and its scalar potential is originating from 
the worldsheet instanton effects in mirror CY. 
Unlike the result of Ref.~\cite{Hebecker:2015rya}, we focus on not only the class of natural inflation, but also 
the more general class of large-field inflation. 
We will show that we can derive new types of inflation potentials, 
which are in a sense a mixture of polynomial functions and sinusoidal functions 
including the natural inflation potential, and 
our new inflation models are favored by the recent results of Planck~\cite{Ade:2015lrj}.

The remaining paper is organized as follows. 
In Sec.~\ref{sec:2}, we review the details of quantum-corrected K\"ahler potential and 
GVW superpotential. 
Based on the moduli stabilization setup as discussed in  Ref.~\cite{Hebecker:2015rya}, 
we show new types of axion inflations and their prediction of cosmological observables in Sec.~\ref{sec:3}. 
The obtained results are favored by the recent results of Planck data and 
the reheating process is different from that of a usual K\"ahler moduli inflation. 
Finally, Sec.~\ref{sec:con} is our conclusion.

\section{Quantum-corrected period vector in type IIB string theory} 
\label{sec:2}

Here and hereafter, we adopt the reduced Planck unit, $M_{\rm Pl}=2.4\times 10^{18}\,{\rm GeV}=1$.
In  type IIB superstring theory on the CY orientifold, all the complex structure moduli ($U$) 
and dilaton ($S$) can be generically stabilized by the flux-induced superpotential~\cite{Giddings:2001yu}, the so-called
GVW superpotential~\cite{Gukov:1999ya},
\begin{align}
W_{\rm GVW}(S,U)=\int G_3(S)\wedge \Omega (U),
\end{align}
where $\Omega$ denotes the holomorphic three-form of the CY manifold and 
$G_3=F_3-i SH_3$ is the linear combination of Ramond-Ramond $(F_3)$ and Neveu-Schwarz three-forms $(H_3)$, respectively. 
Since these three-form fluxes along these cycles 
are quantized on the cycle of the CY manifold, 
GVW superpotential reduces to 
\begin{align}
W_{\rm GVW}(S,U)=\sum_{\alpha=1}^{2h_{-}^{1,2}+2}(N_F-iS N_H)^\alpha \Pi_\alpha,
\label{eq:WGVW}
\end{align}
where $N_F^\alpha, N_H^\alpha$ are integers and $\alpha$ run from $1$ to $2h_{-}^{1,2}+2$ 
with $h_{-}^{1,2}$ being the number of complex structure moduli, i.e., the hodge number of the CY manifold. 
Here, $\Pi_\alpha$ denote the period vector.

In the language of $4$D ${\cal N}=1$ supergravity, 
the K\"ahler potential is described by 
\begin{align}
K=K(U,\bar{U}) -\ln (S+\bar{S}) +K(T,\bar{T}),
\label{eq:Ktot}
\end{align}
where $T$ stand for the K\"ahler moduli. 
Then, 
the scalar potential is expressed by the K\"ahler potential~(\ref{eq:Ktot}) 
and the superpotential~(\ref{eq:WGVW}), 
\begin{align}
V =e^K\left( K^{I\bar{J}} D_IW D_{\bar{J}} \bar{W} -3|W|^2\right),
\end{align}
where $K^{I\bar{J}}$ is the inverse of the K\"ahler metric 
$K_{I\bar{J}}=\partial^2 K/\partial \Phi^I \partial \bar{\Phi}^{\bar{J}}$ for $\Phi^I=S, T^a, U^i$ 
with $a$ being the number of K\"ahler moduli, 
$D_I W=W_I +K_I W$ is the K\"ahler covariant derivative for $W$ with $W_I=\partial W/\partial \Phi^I$, 
and $K_I =\partial K/\partial \Phi^I$. 
From the no-scale structure for the K\"ahler moduli, $K^{a\bar{b}}K_{a}K_{\bar{b}}=3$, 
the scalar potential reduces to 
\begin{align}
V =e^K\left( \sum_{S,U^i} K^{I\bar{J}} D_IW D_{\bar{J}} \bar{W}\right),
\end{align}
from which the complex structure moduli and axion-dilaton are stabilized at the 
supersymmetric minimum, $\langle D_IW\rangle=0$ with $\Phi^I=S, U^i$. 
On the other hand, the K\"ahler moduli would be stabilized by the perturbative string loop and $\alpha^\prime$ 
corrections~\cite{Berg:2005yu} and/or nonperturbative corrections represented by the 
Kachru-Kallosh-Linde-Trivedi (KKLT) scenario~\cite{Kachru:2003aw} or large volume scenario (LVS)~\cite{Balasubramanian:2005zx}.  

Although the above situation is the usual considered setup, one can also consider that 
some of the complex structure moduli would be stabilized through 
not the tree and loop interactions in the GVW superpotential, but the 
worldsheet instanton effects involved in the GVW superpotential. 
In such a case, these light complex structure moduli would be stabilized at their 
minimum after the K\"ahler moduli stabilization. 
It implies that the scale of nonperturbative effects to stabilize the K\"ahler moduli 
is larger than that of worldsheet instanton effects for the period vector. 
Then, one can extract the scalar potential for the light complex structure moduli 
whose real or imaginary parts play a role of inflaton as shown in the next section.

Such quantum corrections for the period vector have been calculated by employing the 
technique of mirror symmetry and ${\cal N}=2$ special geometry in the framework of 
topological string and the gauged linear sigma model on the worldsheet.(See for a review, e.g., Ref.~\cite{Hori:2000kt}.)
The quantum-corrected K\"ahler potential for the complex structure moduli is given by\footnote{See for the case of Calabi-Yau fourfolds~\cite{Honma:2013hma}.}
\begin{align}
e^{-K(U,\bar{U})} &=i \Pi^\dagger\cdot \Sigma \cdot \Pi,\nonumber\\
&=i|X^0|^2\Bigl( 2(F-\bar{F}) -(U^i-\bar{U}^i)(F_i+\bar{F}_i)\Bigl),
\label{eq:Kt}
\end{align}
where $\Sigma$ is the symplectic matrix, 
\begin{align}
\Sigma =
\begin{pmatrix}
0 & \bm{1}\\
-\bm{1} & 0
\end{pmatrix}.
\label{eq:sym}
\end{align}
The prepotential $F$ and its derivative with respect to $X^{\zeta}$ with $\zeta=0,i$ are~\cite{Hosono:1994ax}
\begin{align}
&F=-\frac{1}{3!}\kappa_{ijk}U^iU^jU^k-\frac{1}{2}\kappa_{ij}U^iU^j +\kappa_i U^i +\frac{1}{2}\kappa_0 -\frac{1}{(2\pi i)^3}\sum_\beta n_\beta {\rm Li}_3(q^\beta),
\nonumber\\
&{\cal F}=(X^0)^2 F,\qquad {\cal F}_\zeta=\frac{\partial}{\partial X^\zeta}{\cal F}, \qquad F_i =\frac{\partial}{\partial U^i}F,
\label{eq:pre}
\end{align}
where ${\rm Li}_s(z)=\sum_{n=1}\frac{z^n}{n^s}$ is the polylogarithm function and the integers $n_\beta$ are the genus zero Gromov-Witten invariants
(worldsheet instanton) labeled by $\beta$, i.e., $\beta=d_i \beta_i$ with $d_i$ and $\beta_i$ being the integers and the elements in cohomology $H_2(\tilde{M}_{\rm CY},\bm{Z})\backslash \{0\}$ 
of the mirror CY manifold $\tilde{M}_{\rm CY}$, respectively.   
The explicit form of the period vector $\Pi=(X^0, X^i, {\cal F}_0,{\cal F}_i)^T$ is given by
\begin{align}
\Pi &=X^0
\begin{pmatrix}
1\\
U^i \\
2 F-U^i\partial_i F\\
\partial_i F
\end{pmatrix}
\nonumber\\
&=X^0
\begin{pmatrix}
1\\
U^i \\
\frac{1}{3!}\kappa_{ijk}U^iU^jU^k +\kappa_i U^i +\kappa_0 -\sum_\beta n_\beta^0 (\frac{2}{(2\pi i)^3}{\rm Li}_3(q^\beta) -\frac{d_i}{(2\pi i)^2}U^i{\rm Li}_2(q^\beta))\\
-\frac{1}{2}\kappa_{ijk}U^jU^k -\kappa_{ij}U^j +\kappa_i -\frac{1}{(2\pi i)^2}\sum_\beta n_\beta^0 d_i {\rm Li}_2(q^\beta)
\end{pmatrix}
,
\label{eq:period}
\end{align}
where $\kappa_{ijk}$ is the intersection number of $\tilde{M}_{\rm CY}$, $q^{\beta_i}=e^{2\pi id_iU^i}$ and
\begin{align}
&\kappa_{ijk}=\int_{\tilde{M}_{\rm CY}} J_i \wedge J_j \wedge J_k, \qquad 
\kappa_{ij} =-\frac{1}{2}\int_{\tilde{M}_{\rm CY}} J_i \wedge J_j \wedge J_j, \nonumber\\
&\kappa_j =\frac{1}{4\,3!}\int_{\tilde{M}_{\rm CY}} c_2(\tilde{M}_{\rm CY}) \wedge J_j, \qquad 
\kappa_0 =\frac{\zeta(3)}{(2\pi i)^3}\int_{\tilde{M}_{\rm CY}} c_3(\tilde{M}_{\rm CY}) =\frac{\zeta(3)}{(2\pi i)^3}\chi (\tilde{M}_{\rm CY}).
\end{align}
Here, $\chi (\tilde{M}_{\rm CY})$ is the Euler characteristic and $\zeta(3)\simeq 1.2$. 

By substituting Eqs.~(\ref{eq:sym}), (\ref{eq:pre}) and (\ref{eq:period}) to (\ref{eq:Kt}), the quantum-corrected K\"ahler potential 
is brought into the following form
\begin{align}
e^{-K(U,\bar{U})} &=\frac{i}{6}\sum_{ijk} \kappa_{ijk} (U^i -\bar{U}^i)(U^j -\bar{U}^j)(U^k -\bar{U}^k) 
-\frac{\zeta(3)}{4\pi^3}\chi (\tilde{M}_{\rm CY}) \nonumber\\
&+\frac{i}{(2\pi i)^2} \sum_\beta d_i n_\beta (U^i-\bar{U}^i) \Bigl( {\rm Li}_2(q^\beta)+{\rm Li}_2(\bar{q}^\beta)\Bigl) 
\nonumber\\
&-\frac{2i}{(2\pi i)^3} \sum_\beta n_\beta\Bigl( {\rm Li}_3(q^\beta)+{\rm Li}_3(\bar{q}^\beta)\Bigl),
\label{eq:KT}
\end{align}
where $X^0$ is chosen as unity.
The  term proportional to the Euler characteristic is the $(\alpha^\prime)^3$ correction descended from the 
${\cal R}^4$ term in the ten-dimensional action~\cite{Green:1998by}.
The third and fourth terms in Eq.~(\ref{eq:KT}) denote the worldsheet instanton effects which break the 
continuous shift symmetry to the discrete one.  

In the following, we change the normalization for the complex structure moduli $U^i$ as $iU^i$ and correspondingly the K\"ahler potential is rewritten as
\begin{align}
e^{-K(U,\bar{U})} &=\frac{1}{6}\sum_{ijk} \kappa_{ijk} (U^i +\bar{U}^i)(U^j +\bar{U}^j)(U^k +\bar{U}^k) 
-\frac{\zeta(3)}{4\pi^3}\chi (\tilde{M}_{\rm CY}) \nonumber\\
&+\frac{2}{(2\pi)^2}\sum_\beta \sum_{n=1}^\infty d_i n_\beta (U^i +\bar{U}^i) 
\frac{1}{n^2} {\rm cos}\left( -i\pi n \sum_j d_j (U^j-\bar{U}^j)\right) e^{-\pi n\sum_k d_k (U^k+\bar{U}^k)}
\nonumber\\
&+\frac{4}{(2\pi)^3}\sum_\beta \sum_{n=1}^\infty n_\beta 
\frac{1}{n^3} {\rm cos}\left( -i\pi n \sum_j d_j (U^j-\bar{U}^j)\right) e^{-\pi n\sum_k d_k (U^k+\bar{U}^k)}.
\label{eq:KU}
\end{align}

\section{New type of axion inflation}
\label{sec:3}
In this section, we show the illustrative models where one of the axions of 
complex structure moduli plays a role of inflaton. 
The inflation potential is a mixture of polynomial functions and sinusoidal functions, 
which are descended from the instantonic effects for the period vector of 
CY manifold. 
We assume that the K\"ahler moduli are also stabilized at the KKLT (LVS) minimum in 
Secs.~\ref{subsubsec:KKLT} and~\ref{subsubsec:LVS}, respectively.

\subsection{Moduli stabilization} 

In the following, we denote the $(n-2)$ complex structure moduli 
and axion-dilaton by $z$, whereas the other complex structure moduli 
are represented as $U_1$ and $U_2$. 
By choosing certain three-form fluxes~(\ref{eq:WGVW}) in the 
large complex structure limit~($|z|, |U_1|, |U_2|\gg 1$), 
their K\"ahler potential and superpotential are extracted as
\begin{align}
K({\rm Re}\,S,{\rm Re}\,U)&=-\ln \biggl[f_0\left({\rm Re}\,z, {\rm Re}\,U_1, {\rm Re}\,U_2\right) \biggl],
\nonumber\\
W(S,U) &=g_0(z)+g_1(z)(U_2+N U_1),
\label{eq:KWr}
\end{align}
where $f_{0}$ denotes the first line of Eq.~(\ref{eq:KU}) determined by the topological 
number of the CY manifold, whereas $g_{0,1}(z)$ depends on the quanta of 
three-form fluxes and $N$ is the integer. 
Note that the GVW superpotential is a fourth-order polynomial function at the tree level. 
In analogy of Ref.~\cite{Hebecker:2015rya}, when we redefine the complex structure moduli as 
\begin{align}
&\Psi\equiv U_2+N U_1,
\nonumber\\
&\Phi\equiv U_2,
\end{align}
the K\"ahler potential and superpotential are also rewritten as 
\begin{align}
K({\rm Re}\,S,{\rm Re}\,U)&=-\ln \biggl[f_0\left({\rm Re}\,z, ({\rm Re}\,\Psi-{\rm Re}\,\Phi)/N, {\rm Re}\,\Phi\right) \biggl],
\nonumber\\
W(S,U) &=g_0(z)+g_1(z)\Psi, 
\label{eq:tKW}
\end{align}
which implies that $\Psi$ and $z$ would be stabilized at the 
supersymmetric minimum satisfying $D_{I}W=0$ with $I=\Psi, z$. Correspondingly, the real 
part of $\Phi$ can be stabilized at the supersymmetric minimum at $K_{\Phi}=0$, where 
its mass is given by the nonvanishing superpotential. 
Then, the imaginary part of $\Phi$ remains massless at this stage and ${\rm Im}\,\Phi$ becomes the inflaton 
whose potential is generated by the instanton effects.  

Before going to the detail of instanton effects for ${\rm Im}\,\Phi$, we comment on 
the moduli stabilization for the K\"ahler moduli. 
So far, ${\rm Re}\,\Phi, z, \Psi$ fields are stabilized at the perturbative level. 
When the nonperturbative effects for the K\"ahler moduli appear in 
the superpotential, the K\"ahler moduli can be stabilized 
at the scale above the energy scale of the remaining complex structure modulus ${\rm Im}\,\Phi$. 
As shown later, the mechanism to stabilize the K\"ahler moduli is irrelevant to the 
dynamics of the inflation whose inflaton corresponds to one of the complex 
structure moduli, 
if all the K\"ahler moduli are stabilized at the scale above the energy scale of 
remaining complex structure moduli $\Phi$. 
Thus, one can consider the KKLT scenario or LVS for the K\"ahler 
moduli stabilization with some uplifting scenario. 
In the case of the KKLT scenario, $w=\langle W(S,U)\rangle$ is sufficiently small in order 
to be compatible with the scale of nonperturbative effects for the K\"ahler moduli, whereas 
 $w$ is of order unity in the LVS. 

In certain CY manifold, the instanton effects are explicitly calculated by employing 
the mirror symmetry. 
In the large complex structure moduli limit, one can neglect the exponential term 
in the K\"ahler potential and superpotential due to the large-field values of complex structure moduli 
$z, \Phi, \Psi$. 
However, if we include these corrections, they are represented as 
\begin{align}
\Delta K&\simeq -\frac{1}{\langle f_0\rangle} \sum_{i=1}^2 f_1^{(i)} \left( \frac{2}{\pi}+(U_i+\bar{U}_i)\right) {\rm cos}(-i\pi (U_i-\bar{U}_i))e^{-\pi (U_i+\bar{U}_i)}, 
\nonumber\\
\Delta W &\simeq \sum_{i=1}^2 (g_2^{(i)} +g_3^{(i)} U_i)e^{-2\pi U_i},
\end{align}
where we choose the integer $d_{i}=1$ and the heavy moduli-dependent parameters $g_{2,3}^{(i)}, f_{1}^{(i)}$ are now taken as real, for simplicity. 
The corrections for the K\"ahler potential are extracted in the large complex structure limit. 
Along the line of Ref.~\cite{Hebecker:2015rya}, we omit the exponential term for $U_2$ under 
\begin{align}
e^{-2\pi \langle {\rm Re}U_2\rangle} \ll e^{-2\pi \langle{\rm Re}\, U_1\rangle } \ll 1.
\end{align}
Then, the K\"ahler potential  and superpotential reduce to
\begin{align}
\Delta K&\simeq -\frac{f_1^{(1)}}{\langle f_0\rangle} 
\left( \frac{2}{\pi}+\frac{\Psi+\bar{\Psi}-\Phi-\bar{\Phi}}{N}\right) 
{\rm cos}\left(-i\pi \frac{\Psi -\bar{\Psi}- \Phi+\bar{\Phi}}{N}\right) 
e^{-\pi \frac{\Psi+\bar{\Psi}-\Phi-\bar{\Phi}}{N} }, 
\nonumber\\
\Delta W &\simeq \left(g_2^{(1)} +\frac{g_3^{(1)}}{N}(\Psi -\Phi) \right) 
e^{-2\pi \frac{\Psi-\Phi}{N}},
\label{eq:qKW}
\end{align}
on the basis ($\Psi, \Phi$).

We assume that the other complex structure moduli are enough heavier than the imaginary 
part of $\Phi$ and they are fixed at their supersymmetric minimum. 
Furthermore, the K\"ahler moduli are also fixed at their minimum during and after the inflation. 
This assumption is ensured as follows. 
In our setup, the inflation occurs due to the nonvanishing $\Delta W$ which is mainly determined 
by the real parts of ($\langle \Psi \rangle, \langle \Phi \rangle$). 
Thus, one can choose $\Delta W$ to be smaller 
than the tree-level part of GVW superpotential $w=\langle W\rangle$ and nonperturbative 
superpotential for the K\"ahler moduli. 
In addition, we assume that the nonperturbative effects relevant for the K\"ahler moduli have to be 
independent to the remaining light complex structure modulus $\Phi$, 
otherwise one have to consider the stabilization of K\"ahler moduli and $\Phi$, simultaneously. 
After the stabilization of heavy complex structure moduli and axion-dilaton, 
it is then found that the superpotential behaves effectively constant $w$ 
and K\"ahler moduli can be stabilized 
at the KKLT or LVS without depending on the dynamics of the lightest complex structure modulus, 
as far as the K\"ahler moduli are enough heavier than the lightest modulus.

\subsubsection{Inflaton potential based on KKLT scenario} 
\label{subsubsec:KKLT}

First of all, we assume that the K\"ahler moduli are stabilized at the supersymmetric minimum satisfying 
$D_{T}W=0$ based on the KKLT scenario, 
where the superpotential is effectively given by $W=w+W_{\rm non}(\langle T\rangle)\equiv w_0$ 
with $W_{\rm non}(T)$ being certain nonperturbative effects such as gaugino condensation~\cite{Ferrara:1982qs} 
and D-brane instanton effects.\footnote{The backreaction from the K\"ahler moduli is left for future work.} 
It is then further assumed that the prefactor of nonperturbative 
effects for the K\"ahler moduli do not depend on $\Phi$. 
In such a case, when the gauginos living on $N$ stacks of D$7$-branes condensate, 
the nonperturbative superpotential $W_{\rm non}(T)$ is described as $W_{\rm non}(T)\simeq e^{-aT}$ with $a=8\pi^2/N$. 
The mass scales of K\"ahler moduli, $m_T^2\simeq (a\,m_{3/2})^2$ with 
$m_{3/2}\simeq e^{\left(K({\rm Re}\,S,{\rm Re}\,U)+K(T,\bar{T})\right)/2} w$, are much heavier than the remaining complex structure modulus which is determined 
by $\Delta W$.

Next, we now turn to the dynamics of the lightest complex structure modulus ${\rm Im}\,\Phi$. 
With the tree level and quantum-corrected K\"ahler potential and superpotential~(\ref{eq:tKW},\ref{eq:qKW}), 
the superpotential and its covariant derivative become 
\begin{align}
W&\simeq w_0+\Delta W,
\nonumber\\
D_{\Phi}W&\simeq \Delta W_{\Phi}+\Delta K_{\Phi}W
\nonumber\\
&\simeq e^{-2\pi \langle {\rm Re}\,U_1\rangle }  \biggl[ 
e^{-2\pi i\frac{\langle {\rm Im}\,\Psi\rangle -{\rm Im}\,\Phi}{N}} \left(
-\frac{g_3^{(1)}}{N}+\frac{2\pi}{N} \left(g_2^{(1)} +\frac{g_3^{(1)}}{N}(\langle \Psi \rangle -\Phi)\right) 
\right)
\nonumber\\
&
-\frac{w_0f_1^{(1)}}{\langle f_0\rangle} \left(
-\frac{1}{N}{\rm cos}\left(2\pi  \frac{\langle{\rm Im}\,\Psi \rangle- {\rm Im}\,\Phi}{N}\right)  
+\frac{\pi}{N}\left( \frac{2}{\pi}+2\langle {\rm Re}\,U_1 \rangle\right) 
e^{-2\pi i\frac{\langle {\rm Im}\,\Psi\rangle -{\rm Im}\,\Phi}{N}} 
\right)
\biggl], 
\end{align}
in the limit of $w_0 \gg \Delta W$. 
\clearpage
These lead to the scalar potential for $\tilde{\phi}={\rm Im}\,\Phi$,\footnote{In the framework of LVS, 
the $-3|W|^2$~term in the scalar potential is canceled by the extended no-scale structure of K\"ahler moduli~\cite{Cicoli:2007xp} as shown in Sec.~\ref{subsubsec:LVS}.}
\begin{align}
e^{-K-\Delta K}V&\simeq K^{\Phi\bar{\Phi}} D_{\Phi}WD_{\bar{\Phi}}\bar{W}-3|W|^2
\nonumber\\
&\simeq -3| W|^2 +K^{\Phi \bar{\Phi}}  e^{-4\pi \langle{\rm Re}\,U_1\rangle}
\biggl[ A^2+B^2 (\langle{\rm Im}\,\Psi \rangle -\tilde{\phi})^2  
\nonumber\\
&+(C^2 +2AC){\rm cos}^2\left(2\pi  \frac{\langle{\rm Im}\,\Psi \rangle- \tilde{\phi}}{N}\right) 
+BC (\langle{\rm Im}\,\Psi \rangle- \tilde{\phi})\, {\rm sin}\left(4\pi  \frac{\langle{\rm Im}\,\Psi \rangle- \tilde{\phi}}{N}\right)
\biggl] 
\nonumber\\
&\simeq -3w_0^2 -6w_0e^{-2\pi \langle{\rm Re}\,U_1\rangle}
\biggl[ \left(g_2^{(1)}+\frac{g_3^{(1)}}{N}\langle {\rm Re}\,U_1\rangle  \right) 
{\rm cos}\left(2\pi  \frac{\langle{\rm Im}\,\Psi \rangle- \tilde{\phi}}{N}\right) 
\nonumber\\
&
+\frac{g_3^{(1)}}{N} (\langle{\rm Im}\,\Psi \rangle- \tilde{\phi}) 
{\rm sin}\left(2\pi  \frac{\langle{\rm Im}\,\Psi \rangle- \tilde{\phi}}{N}\right) 
\biggl] 
\nonumber\\
&-3\,e^{-4\pi \langle{ \rm Re}\,U_1\rangle}\biggl[ 
\left(g_2^{(1)}+\frac{g_3^{(1)}}{N}\langle {\rm Re}\,U_1\rangle \right)^2
+\left(\frac{g_3^{(1)}}{N}\right)^2 (\langle {\rm Im}\,\Psi \rangle -\tilde{\phi}  )^2\biggl]
\nonumber\\
&+K^{\Phi \bar{\Phi}}  e^{-4\pi \langle{\rm Re}\,U_1 \rangle}
\biggl[A^2+B^2 (\langle {\rm Im}\,\Psi\rangle -\tilde{\phi})^2  
\nonumber\\
&+(C^2 +2AC){\rm cos}^2\left(2\pi  \frac{\langle{\rm Im}\,\Psi \rangle- \tilde{\phi}}{N}\right) 
+BC (\langle{\rm Im}\,\Psi \rangle-\tilde{\phi})\, {\rm sin}\left(4\pi  \frac{\langle{\rm Im}\,\Psi \rangle- \tilde{\phi}}{N}\right)
\biggl]  ,
\label{eq:scalar}
\end{align}
where $K=K({\rm Re}\,S,{\rm Re}\,U)+\Delta K +K(T,{\bar T})$ is the full K\"ahler potential of 
moduli fields. The other parameters are defined as
\begin{align}
A&=-\frac{g_3^{(1)}}{N}+\frac{2\pi g_2^{(1)}}{N} +\frac{2\pi g_3^{(1)}}{N} \langle {\rm Re}\,U_1 \rangle -\frac{w_0f_1^{(1)}2\pi}{\langle f_0\rangle N} \left(\frac{1}{\pi}+\langle {\rm Re}\,U_1 \rangle\right),
\nonumber\\
B&=\frac{2\pi g_3^{(1)}}{N^2},\,\,\,\, C=\frac{w_0f_1^{(1)}}{\langle f_0\rangle N }.
\end{align}
When the complex structure moduli are canonically normalized at the Minkowski minimum by setting 
 certain uplifting such as the anti-D-brane~\cite{Kachru:2003aw} and F-term uplifting scenario~\cite{Lebedev:2006qq,Dudas:2006gr}, 
the inflaton potential is approximately given in the large complex structure 
limit $\langle{\rm Re}\,U_1\rangle \gg 1$,
\begin{align}
V_{\rm inf}&\simeq \Lambda_1\left( 1-{\rm cos}\frac{\phi}{M_1} \right)
+\Lambda_2 \phi \,{\rm sin}\frac{ \phi}{M_1}, 
\label{eq:inf1}
\end{align}
where $\phi\equiv  k_1(\langle{\rm Im}\,\Psi \rangle -\tilde{\phi})$ with $k_1$ being the normalization 
factor of ${\cal O}(\langle{\rm Re}\,U_2\rangle)$; $M_1=N k_1/2\pi$ and 
$\Lambda_{1,2}\simeq {\cal O}(e^{\langle K\rangle}w_0e^{-2\pi \langle{\rm Re}\,U_1\rangle})$ 
are the real constants. 
The first term is the same as the potential form of natural inflation, but the 
second term as well as this full potential is new.
Note that although this potential form looks slightly similar to the natural inflation 
with modulation\footnote{See e.g., Refs.~\cite{Kobayashi:2010pz,Czerny:2014wua,Kobayashi:2014ooa}.}, 
the second term is not a small correction, but two terms are comparable 
with each other.

So far, we have neglected the higher order terms of the worldsheet instanton effects, because the constant $w_0$ 
dominates the superpotential. 
When the K\"ahler moduli are stabilized at the minimum suggested by Kallosh-Linde~\cite{Kallosh:2004yh}, 
the constant $w_0$ can be taken as $w_0 \sim 0$ due to the cancelation between the nonperturbative effects for the K\"ahler moduli $W_{\rm non}(\langle T\rangle)$ and the part of the tree-level GVW superpotential $w$. 
The mass scales of K\"ahler moduli, $m_{T}^2 \sim w^2$, are also much heavier than that of the remaining 
complex structure modulus which is determined by the higher order terms of the worldsheet instanton effects. 
In such a case, the inflaton potential is dominated by the order of $e^{-4\pi \langle{\rm Re}\,U_1\rangle}$ 
in the scalar potential~(\ref{eq:scalar}),
\begin{align}
V_{\rm inf}&\simeq \Lambda_3 \phi^2,
\label{eq:inf2}
\end{align}
where $\phi\equiv  k_2(\langle{\rm Im}\,\Psi \rangle -\tilde{\phi})$ with $k_2$ being the normalization 
factor  of ${\cal O}(\langle{\rm Re}\,U_2\rangle)$ and 
$\Lambda_{3}\simeq {\cal O}(e^{\langle K\rangle} e^{-4\pi \langle{\rm Re}\, U_1\rangle})$ are 
constants by setting a specific uplifting sector in the same way as Eq.~(\ref{eq:inf1}). 
However, in this situation, we need to tune the parameter in the sector of K\"ahler moduli so that 
the mass scale of ${\rm Re}\,\Phi$ is larger than that of  inflaton. 
The obtained scalar potential is just that of chaotic inflation.

\subsubsection{Inflaton potential based on the large volume scenario} 
\label{subsubsec:LVS}

Next, we assume that the K\"ahler moduli are stabilized at the nonsupersymmetric 
minimum within the framework of LVS, where the $\alpha^\prime$ corrections and 
nonperturbative effects are included in the K\"ahler potential and 
superpotential, respectively. 

In the same way as the case of Sec.~\ref{subsubsec:KKLT}, we consider that the K\"ahler moduli 
are enough heavier than the remaining complex structure modulus ${\rm Im}\,\Phi$. 
Because of the extended no-scale structure of the K\"ahler moduli~\cite{Cicoli:2007xp}, 
the $-3|W|^2$ term in the scalar potential is vanishing and Eq.~(\ref{eq:scalar}) reduces to
\begin{align}
V&\simeq e^{K+\Delta K}K^{\Phi\bar{\Phi}} D_{\Phi}WD_{\bar{\Phi}}\bar{W}
\nonumber\\
&\simeq e^{K+\Delta K}K^{\Phi\bar{\Phi}}  e^{-4\pi \langle{\rm Re}\,U_1\rangle}
\biggl[ A^2+B^2 ({\rm Im}\,\Psi-\tilde{\phi})^2  
\nonumber\\
&+(C^2 +2AC){\rm cos}^2\left(2\pi  \frac{\langle{\rm Im}\,\Psi \rangle- \tilde{\phi}}{N}\right) 
+BC (\langle{\rm Im}\,\Psi \rangle- \tilde{\phi})\, {\rm sin}\left(4\pi  \frac{\langle{\rm Im}\,\Psi \rangle- \tilde{\phi}}{N}\right)
\biggl] .
\label{eq:scalar2}
\end{align}
This leads to the following scalar potential for the canonically normalized field 
$\phi\equiv k_3\left( \langle{\rm Im}\,\Psi\rangle -\tilde{\phi}\right)$ with $k_3$ being the normalization factor 
 of ${\cal O}(\langle{\rm Re}\,U_2\rangle)$:
\begin{align}
V_{\rm inf}&\simeq \Lambda_4 \phi^2 +\Lambda_5 \phi\, 
{\rm sin}\left(\frac{\phi}{M_3}\right) 
+\Lambda_6\left(1-{\rm cos}\left(\frac{\phi}{M_3}\right) \right),
\label{eq:inf3}
\end{align}
where $M_3=N k_3/4\pi$, $\Lambda_{4,5,6}\simeq {\cal O}(e^{\langle K\rangle} e^{-4\pi \langle{\rm Re}\, U_1\rangle})$ 
are constants. Here we set some uplifting sector to obtain the Minkowski minimum. 
Thus, this potential is also new and a mixture of polynomial functions and sinusoidal functions.
These three terms are comparable with each other. 

Although we have neglected a backreaction from the K\"ahler moduli, 
its scalar potential is roughly estimated as
\begin{align}
V_{\rm LVS}&\sim e^{K(S,U)} \frac{|W|^2}{{\cal V}^3},
\label{eq:LVS}
\end{align}
where ${\cal V}$ denotes the volume of the CY manifold. 
From  Eq.~(\ref{eq:LVS}), the scalar potential for $\phi$ appearing from Eq.~(\ref{eq:LVS}) 
is of order ${\cal O}(e^{\langle K(S,U)\rangle} w_0e^{-2\pi \langle {\rm Re}\,U_1\rangle}/{\cal V}^3)$. 
Thus, under 
\begin{align}
e^{\langle K(S,U)\rangle} \frac{w_0e^{-2\pi \langle {\rm Re}\,U_1\rangle}}{{\cal V}^3} < 
e^{\langle K(S,U)\rangle} \frac{e^{-4\pi \langle{\rm Re}\, U_1\rangle}}{{\cal V}^2},
\end{align}
the potential arising from the backreaction of K\"ahler moduli can be negligible. 
This situation can be realized with the mild large volume of the CY manifold, e.g., 
${\cal V}\simeq 10^2$, ${\rm Re}\,U_1\simeq 0.7$, and $w_0\simeq 1$. 
When we include the backreaction from the K\"ahler moduli, 
it induces the scalar potential as in Eq.~(\ref{eq:inf1}). 
Then, the total scalar potential is described by that of Eq.~(\ref{eq:inf3}) with 
$\Lambda_{5,6}>\Lambda_4$. 
We leave the detailed study of the backreaction to future work.
Note that the energy scale of K\"ahler moduli $V_{\rm LVS}$ should be also larger than 
the inflation scale $V_{\rm inf}$, i.e., $V_{\rm inf} \ll V_{\rm LVS}$.

\subsection{Numerical analyses}
\label{sub:num}

\begin{figure}[ht]
\centering \leavevmode
\includegraphics[width=0.9\linewidth]{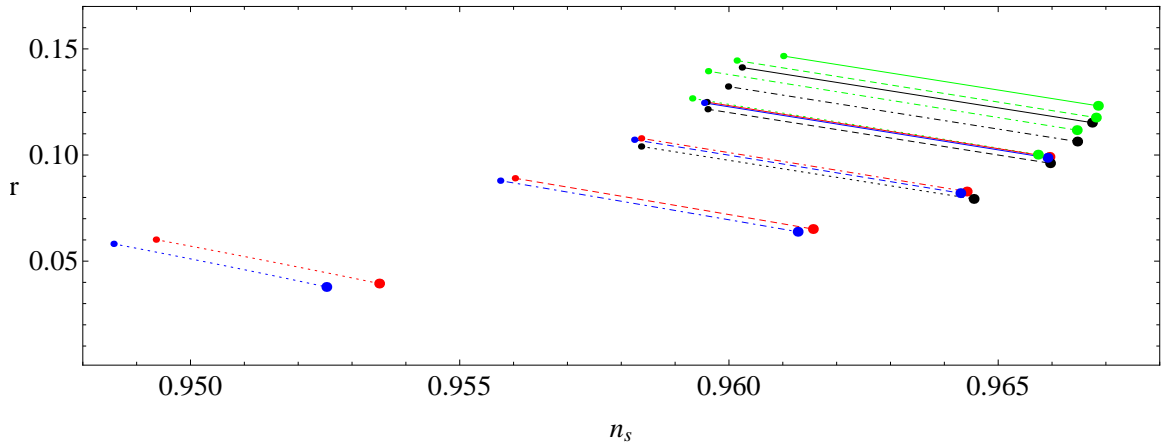}
\caption{Predictions of $(n_s, r)$ in the range of 
{\it e}-folding number, $50\leq N_e \leq 60$ for the inflaton potential~(\ref{eq:inf1}) 
with and without sinusoidal term. 
The leftmost (rightmost) circle on each line represents for the {\it e}-folding number $N_e=50$ ($N_e=60$).  
The black line corresponds to the prediction of natural inflation. 
For the red, blue and green lines, the sinusoidal term is 
included as $\Lambda_2/\Lambda_1=1, 5, -1/6M_1$, respectively. 
The solid, dot-dashed, dashed, and dotted lines denote the 
axion decay constants $M_1=15,12, 10, 8$, respectively.}
\label{fig:inst1}
\end{figure}

In this section, we study the inflaton dynamics characterized by three types of 
inflaton potential~(\ref{eq:inf1}), (\ref{eq:inf2}), and (\ref{eq:inf3}). 
First of all, the inflaton potential in Eq.~(\ref{eq:inf2}) is just the form of chaotic inflation which 
is disfavored by the recent result of Planck. 
In addition to it, the inflaton potential (\ref{eq:inf1}) is that of natural inflation with the sinusoidal 
term. In our setup, as discussed in Ref.~\cite{Hebecker:2015rya}, the axion decay constant 
takes the value of the trans-Planckian due to the nature of winding trajectory on the ($\Psi, \Phi$) plane, 
i.e., $M_2\gg 1$. 
Thus, the scalar potential of natural inflation without the sinusoidal term~(\ref{eq:inf1}) 
is agreement with the Planck data~\cite{Ade:2015lrj} under its large axion decay constant, 
\begin{align} 
P_\xi&= 2.20 \pm 0.10 \times 10^{-9},
\nonumber\\
n_s&= 0.9655\pm 0.0062,
\nonumber\\
r&<0.11,
\end{align} 
where $P_\xi$ is the power spectrum of curvature perturbation and 
$r$ denotes the tensor-to-scalar ratio at the pivot scale $k_{\ast}=0.05\,{\rm Mpc}^{-1}$. 
In what follows, we choose that the inflation scale is set to 
realize the correct magnitude of the power spectrum of curvature perturbation. 
For the inflaton potential (\ref{eq:inf1}), the slow-roll parameters are 
estimated as
\begin{align} 
\epsilon &=\frac{1}{2}\left(\frac{\partial_\phi V}{V}\right)^2 
=\frac{1}{2}\left(\frac{\left(\frac{\Lambda_1}{M_1}+\Lambda_2\right){\rm sin}\frac{\phi}{M_1} 
+\Lambda_2\frac{\phi}{M_1}{\rm cos}\frac{\phi}{M_1}}
{\Lambda_1 \left(1-{\rm cos}\frac{\phi}{M_1}\right) +\Lambda_2\phi\, {\rm sin}\frac{\phi}{M_1}}\right)^2
\simeq \frac{2}{\phi^2}-\frac{1}{3M_1^2} 
\frac{\Lambda_1+4M_1\Lambda_2}{\Lambda_1+2M_1\Lambda_2},
\nonumber\\
\eta &=\frac{\partial_\phi \partial_\phi V}{V}
=\frac{\frac{1}{M_1}\left(\frac{\Lambda_1}{M_1}+2\Lambda_2\right){\rm cos}\frac{\phi}{M_1} 
-\Lambda_2\frac{\phi}{M_1^2}{\rm sin}\frac{\phi}{M_1}}
{\Lambda_1 \left(1-{\rm cos}\frac{\phi}{M_1}\right) +\Lambda_2\phi\, {\rm sin}\frac{\phi}{M_1}}
\simeq \frac{2}{\phi^2}-\frac{5}{6M_1^2} 
\frac{\Lambda_1+4M_1\Lambda_2}{\Lambda_1+2M_1\Lambda_2},
\end{align} 
and the $n_s$ and $r$ become
\begin{align} 
n_s &=1-6\epsilon +2\eta 
\simeq 1-\frac{8}{\phi^2}+\frac{1}{3M_1^2} 
\left(1+
\frac{2M_1\Lambda_2}{\Lambda_1+2M_1\Lambda_2}
\right),
\nonumber\\
r &=16\epsilon \simeq \frac{32}{\phi^2}-\frac{16}{3M_1^2} 
\left( 1+
\frac{2M_1\Lambda_2}{\Lambda_1+2M_1\Lambda_2}
\right).
\label{eq:nsr1}
\end{align} 
It is found that $n_s$ ($r$) tends to be small (large) when $\Lambda_2$ becomes negative. 
We now take into account the condition that the inflaton mass should be positive at least at a vacuum, that is, 
$\partial^2 V/\partial \phi^2=(\Lambda_1+2M_1\Lambda_2)/M_1^2 >0$. 
These features are also captured in Fig.~\ref{fig:inst1}. 
However, as can be seen in Fig.~\ref{fig:inst1}, there are no sizable differences between 
the prediction of $n_s$ and $r$ 
of natural inflation with and without the sinusoidal term. 
In any cases, the large axion decay constant is favored by the Planck data~\cite{Ade:2015lrj} 
in the light of the spectral tilt of curvature perturbation. 
The typical values of cosmological observables associated with a sufficient {\it e}-folding  number 
are summarized in Table~\ref{tab:result1}.  

\begin{figure}[ht]
\centering \leavevmode
\includegraphics[width=0.9\linewidth]{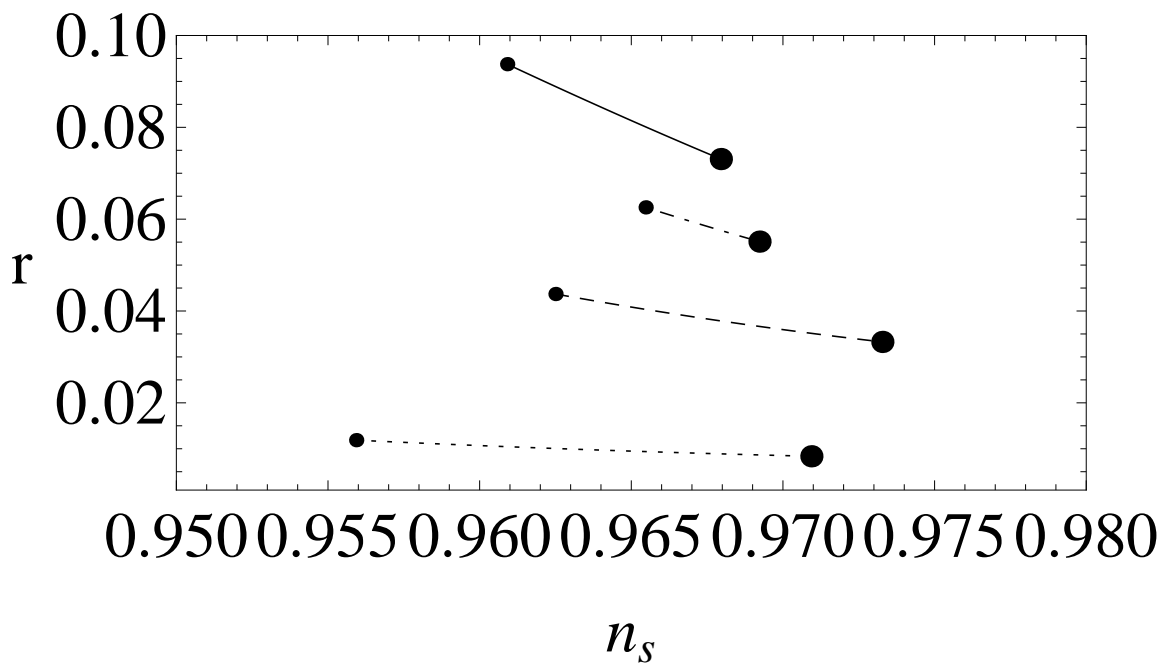}
\caption{Predictions of $(n_s, r)$ in the range of 
{\it e}-folding number, $50\leq N_e \leq 60$ for the inflaton potential~(\ref{eq:inf3}) 
with  $\Lambda_4/\Lambda_6=1$ and  $\Lambda_5/\Lambda_6=5$. 
The leftmost (rightmost) circle on each line represents for the {\it e}-folding number $N_e=50$ ($N_e=60$). 
The black solid, black dot-dashed, black dashed and black dotted 
lines correspond to the axion decay constants 
$M_3=6, 5, 4, 3$, respectively.}
\label{fig:inst2}
\end{figure}

Finally, let us discuss the scalar potential in Eq.~(\ref{eq:inf3}).
The slow-roll parameters are also estimated as 
\begin{align} 
\epsilon &=\frac{1}{2}\left(\frac{2\Lambda_4\phi +(\Lambda_5 +\frac{\Lambda_6}{M_3}){\rm sin}\frac{\phi}{M_3} +\frac{\Lambda_5}{M_3}\phi\,{\rm cos}\frac{\phi}{M_3}}
{\Lambda_4 \phi^2 +\Lambda_5\phi\,{\rm sin}\frac{\phi}{M_3}+\Lambda_6(1-{\rm cos}\frac{\phi}{M_3})}\right)^2
\simeq \frac{2}{\phi^2}-\frac{4M_3\Lambda_5 +\Lambda_6}{3M_3^2\left(2M_3(M_3\Lambda_4+\Lambda_5)+\Lambda_6\right)}+{\cal O}\left(\frac{1}{M_3^4}\right),
\nonumber\\
\eta &=\frac{2\Lambda_4-\frac{\Lambda_5}{M_3^2}\phi\,{\rm sin}\frac{\phi}{M_3} +\frac{1}{M_3}(2\Lambda_5+\frac{\Lambda_6}{M_3}){\rm cos}\frac{\phi}{M_3}}
{\Lambda_4 \phi^2 +\Lambda_5\phi\,{\rm sin}\frac{\phi}{M_3}+\Lambda_6(1-{\rm cos}\frac{\phi}{M_3})}
\simeq \frac{2}{\phi^2}-\frac{5(4M_3\Lambda_5 +\Lambda_6)}{6M_3^2\left(2M_3(M_3\Lambda_4+\Lambda_5)+\Lambda_6\right)}+{\cal O}\left(\frac{1}{M_3^4}\right),
\end{align} 
which leads to the expressions of $n_s$ and $r$,
\begin{align} 
n_s &\simeq 1-\frac{8}{\phi^2}+\frac{4M_3\Lambda_5 +\Lambda_6}{3M_3^2\left(2M_3(M_3\Lambda_4+\Lambda_5)+\Lambda_6\right)}+{\cal O}\left(\frac{1}{M_3^4}\right),
\nonumber\\
r &\simeq \frac{32}{\phi^2}-\frac{16(4M_3\Lambda_5 +\Lambda_6)}{3M_3^2\left(2M_3(M_3\Lambda_4+\Lambda_5)+\Lambda_6\right)}+{\cal O}\left(\frac{1}{M_3^4}\right).
\end{align} 
Then, it is found that 
$n_s$ ($r$) tends to be large (small) when $\Lambda_5$ becomes large compared with $\Lambda_4$ 
in the case of $M_3 \gg 1$. 
We now take into account the condition that the inflaton mass should be positive at least at a vacuum, that is, 
$\partial^2 V/\partial \phi^2=(2M_3(M_3\Lambda_4+\Lambda_5)+\Lambda_6)/M_3 >0$. 
These features are also captured in Fig.~\ref{fig:inst2}. 
With suitable parameters, 
the cosmological observables associated with a sufficient {\it e}-folding number are summarized in Table~\ref{tab:result2} 
which are within the central value of recent Planck data~\cite{Ade:2015lrj}.
From Fig.~\ref{fig:inst2}, it is found that the small- and large-field inflations can be realized 
without changing the spectral tilt of adiabatic curvature perturbation. 
The small-field inflation is achieved by the plateau appearing in the scalar potential.

\begin{table}
\begin{center}
\begin{tabular}{|c|c|c|c|c|c|} \hline 
$M_1$ & $\Lambda_2/\Lambda_1$ &  $N_e$& $n_s$ & $r$ & $dn_s/d\ln k$ 
\\ \hline
$8$ & $5$ & $50$ & $0.96$ & $0.04$ & $-0.0005$ 
\\\hline
$8$ & $5$ &  $60$ & $0.964$ & $0.03$ & $-0.0003$ 
\\\hline
$10$ & $5$  & $50$ & $0.964$ & $0.055$ & $-0.0006$ 
\\\hline
$10$ & $5$  & $60$ & $0.969$ & $0.041$ & $-0.0004$ 
\\\hline
$12$ & $5$  & $50$ & $0.966$ & $0.063$ & $-0.0006$ 
\\\hline
$12$ & $5$  & $60$ & $0.971$ & $0.049$ & $-0.0004$ 
\\\hline
$15$ & $5$ &  $50$ & $0.968$ & $0.07$ & $-0.0006$ 
\\\hline
$15$ & $5$ &  $60$ & $0.973$ & $0.056$ & $-0.0004$ 
\\\hline
 \end{tabular}
 \caption{The input parameters $M_1$, $\Lambda_2/\Lambda_1$ 
and the output values of the {\it e}-folding number $N_e$, 
spectral tilt of curvature perturbation $n_s$, tensor-to-scalar 
ratio $r$ and the running of spectral index $dn_s/d\ln k$.}
\label{tab:result1}
\end{center}
\end{table}

\clearpage

\begin{table}
\begin{center}
\begin{tabular}{|c|c|c|c|c|c|c|} \hline 
$M_3$ & $\Lambda_4/\Lambda_6$ & $\Lambda_5/\Lambda_6$  & $N_e$& $n_s$ & $r$ & $dn_s/d\ln k$ 
\\ \hline
$5$ & $1$ & $5$ & $60$ & $0.969$ & $0.05$ & $-0.0007$ 
\\\hline
$5$ & $1$ & $5$ & $55$ & $0.965$ & $0.06$ & $-0.0008$ 
\\\hline
$5$ & $1$ & $5$ & $50$ & $0.962$ & $0.07$ & $-0.0009$ 
\\\hline
$3$ & $1$ & $5$ & $60$ & $0.97$ & $0.008$ & $0.0009$ 
\\\hline
$3$ & $1$ & $5$ & $55$ & $0.964$ & $0.097$ & $0.0009$ 
\\\hline
$3$ & $1$ & $5$ & $50$ & $0.956$ & $0.012$ & $0.0009$ 
\\\hline
$5$ & $1/5$ & $1$ & $60$ & $0.968$ & $0.05$ & $-0.0007$ 
\\\hline
$5$ & $1/5$ & $1$ & $55$ & $0.965$ & $0.06$ & $-0.0008$ 
\\\hline
 \end{tabular}
 \caption{The input parameters $M_3$, $\Lambda_4/\Lambda_6$, 
$\Lambda_5/\Lambda_6$ and the 
output values of the {\it e}-folding number $N_e$, 
spectral tilt of curvature perturbation $n_s$, tensor-to-scalar 
ratio $r$, and the running of spectral index $dn_s/d\ln k$.}
\label{tab:result2}
\end{center}
\end{table}

\clearpage

\subsection{Reheating temperature}

In the previous numerical analyses, we roughly estimate the 
amount of {\it e}-folding within the range $50\leq N_e \leq 60$. 
However, it is sensitive to the thermal history of the Universe 
after the inflation. 
Indeed, the reheating temperature gives the significant 
effect as can be seen from the simple formula of the amount of 
{\it e}-folding~\cite{Liddle:1993fq}
\begin{align}
N_e \simeq 62 + 
\text{ln} \frac{V_{\ast}^{1/4}}{10^{16}\,{\rm GeV}} 
+\text{ln} \frac{V_\ast^{1/4}}{V_{\rm end}^{1/4}} 
-\frac{1}{3} \text{ln} \frac{V_{\rm end}^{1/4}}{\rho_R^{1/4}},
\label{eq:efold}
\end{align}
where $V_\ast^{1/4}$ ($V_{\rm end}^{1/4}$) is the 
energy density of scalar potential at the horizon exit (inflation end), 
and $\rho_R^{1/4} =(\pi^2 g_\ast/30)T_R$ is the energy density of radiation, 
and  $g_\ast$ is 
the effective degrees of freedom of the radiation at the reheating 
temperature $T_R$.

In this section, we show the typical reheating temperature when one of the 
complex structure moduli plays a role of inflaton in three types of inflaton 
potential~(\ref{eq:inf1})-(\ref{eq:inf3}). 
After the inflation, the energy of inflaton is converted into that of matter fields 
in the visible and hidden sectors through the inflaton decay into them. 
In the string inflation with K\"ahler moduli, the inflaton  mainly 
decays into the gauge bosons, gauginos, or Higgs fields due to a dimensional 
counting.(See for more details, e.g., Ref.~\cite{Higaki:2012ba} in the case of LVS.) 
However, the complex structure moduli 
do not appear in the gauge kinetic function at the tree level 
unlike the case of the K\"ahler moduli. 
In the following analysis, we assume that the supersymmetry breaking is caused by the 
K\"ahler moduli or matter fields at a scale larger than the inflation scale. 
Therefore, below the inflation scale, the effective theory would be described by 
that of the standard model. 
The following mass formula of Weyl fermions: 
\begin{align}
m_{IJ}=e^{G/2}\left( \nabla_I G_{J} +\frac{1}{3}G_I G_J \right),
\end{align}
with $\nabla_I G_J =\partial G_J/\partial \Phi^I -\Gamma^K_{IJ}G_K$, $G=K+\ln (\bar{W}W)$, 
$G_I=\partial G/\partial \Phi^I$, and 
$\Gamma^K_{IJ}$ is the Christoffel symbol constructed by the metric $K_{IJ}$, 
leads to the mass of the Weyl fermion paired with the inflaton field ($\psi$)
\begin{align}
m_{\psi \psi}\simeq e^{G/2} G_{\Phi \bar{\Phi}},
\end{align}
which is almost of the same order as the supersymmetry breaking scale due to $G_{\Phi}\simeq 0$. 
When we neglect the inflaton decay into $\psi$, 
the leading decay channels are categorized into the following three types of them.

First of all, the inflaton decays into the gauge bosons in the standard model 
through the moduli-dependent one-loop corrections in the gauge kinetic function~\cite{Dixon:1990pc},
\begin{align}
{\cal L}=-\frac{1}{4g_a^2}F^a_{\mu\nu}F^{a\mu\nu}-\frac{1}{4}\frac{\Delta(\Phi)}{16\pi^2} F^a_{\mu\nu}F^{a\mu\nu}
\end{align}
where $\Delta (\Phi)$ is unknown threshold correction for $\Phi$\footnote{The explicit form 
of it on the toroidal background can be shown in Ref.~\cite{Lust:2003ky}.} 
and $a=1,2,3$ denote the gauge groups in the standard model, $U(1)_Y$, 
$SU(2)_L$, and $SU(3)_C$, respectively. 
Thus, its decay width  is estimated as 
\begin{align}
\Gamma_\phi^{(1)} \equiv  \sum_{a=1}^3 \Gamma (\phi \rightarrow g^{(a)}+g^{(a)})
=\sum_{a=1}^3\frac{N_G^a}{128\pi}\left(\frac{\partial_\phi (\Delta(\Phi)) g_a^2}{16\pi^2 d} \right)^2 
\frac{m_\phi^3}{M_{\rm Pl}^2} 
\simeq 5.8 \times 10^{-5}\left(\frac{m_\phi}{10^{13}\,{\rm GeV}}\right)^3\,{\rm GeV},
\end{align}
where $\sum_aN_G^a=12$, $d \simeq {\cal O}(\sqrt{K_{\Phi \bar{\Phi}}}) \simeq {\cal O}(1)$, 
$m_\phi$ is the inflaton mass,  
and $(g_a)^{2}\simeq 0.53$ is the gauge coupling at the grand unification 
scale~$2\times 10^{16}\, {\rm GeV}$.\footnote{
Although we implicitly assume that the minimal supersymmetric standard model can be realized 
at the grand unification scale, the size of gauge coupling is not relevant to our following analysis.} 
We now assume that the threshold correction is provided by the linear term of $\Phi$ in the large 
complex structure limit, that is, $\Delta (\Phi)=\Phi$ as can be also seen in the toroidal background~\cite{Lust:2003ky}. 
When the total decay width of inflaton is dominated by the above decay, 
the reheating temperature is roughly estimated by 
equaling the Hubble parameter at the reheating and the total decay width, 
\begin{align}
\Gamma_\phi^{(1)} &\simeq H (T_R), \nonumber\\
\Rightarrow T_{R} &= 
\left( \cfrac{\pi^2 g_\ast}{90}\right)^{-1/4} 
\sqrt{\Gamma_\phi^{(1)} M_{\rm Pl}} \simeq  6.4\times 10^6 
\left(\frac{m_\phi}{10^{13}\,{\rm GeV}}\right)^{3/2}
\,{\rm GeV},
\end{align}
where $g_\ast =106.75$ is the effective degrees of 
freedom of the radiation at the reheating in the standard model. 
The inflaton decay into the gaugino pairs is prohibited 
when the gauginos are heavier than the inflaton. 
Also, anomaly-induced inflaton decays are subdominant effect because of $K_{\Phi}\simeq 0$~\cite{Endo:2007ih}.

The next one is the inflaton decay into the matter fields $Q$ in the standard model. 
Their couplings would appear from the following K\"ahler potential:\footnote{We 
use the same notation between the inflaton $\phi$ and its supermultiplet.}
\begin{align}
K \simeq f(T+\bar{T})e^{-2\pi {\rm Re}\,U_1}{\rm cos}\left(\frac{\phi}{M}\right) |Q|^2,
\end{align}
where $M$ is the axion decay constant of inflaton, 
and $f(T+\bar{T})$ is some K\"ahelr moduli-dependent function.
Such an instantonic coupling is the leading one between the axion inflaton 
and matter fields. This is because the axion does not appear at the tree-level 
K\"ahler potential due to its shift symmetry and at the nonperturbative 
level, the continuous shift symmetry is broken into the discrete one. 
It is then found that the above coupling does not lead to the inflaton decay into 
the matter fermions or bosons, because the supersymmetry is broken at the scale 
larger than the inflation scale. 

Finally, we explore the couplings between the inflaton and matter fields 
in the superpotential, although they are higher order terms. 
The relevant interactions would appear from the Yukawa couplings,
\begin{align}
W=Y_{ijk}(\Phi) Q^iQ^jQ^k,
\end{align}
where $Y_{ijk}(\Phi)$ denotes the physical Yukawa coupling between the 
canonically normalized matter chiral multiplets $Q_i$ in the visible sector (see, e.g., Ref.~\cite{Cremades:2004wa} in 
the case of toroidal background). 
That leads to the following decay width,
\begin{align}
\Gamma_\phi^{(2)} \equiv  \Gamma (\phi \rightarrow \psi^i \psi^j Q^k)
\simeq \frac{1}{64(2\pi)^3}\left\langle \frac{\partial Y_{ijk}}{\partial \phi}\right\rangle^2 \frac{m_\phi^3}{M_{\rm Pl}^2},
\end{align}
where $\psi^i$ and $Q^i$ denote the matter fermion and boson in the chiral multiplet $Q^i$, 
respectively.
The associated reheating temperature is roughly estimated as
\begin{align}
\Gamma_\phi^{(2)} &\simeq H (T_R), \nonumber\\
\Rightarrow T_{R} &= 
\left( \cfrac{\pi^2 g_\ast}{90}\right)^{-1/4} 
\sqrt{\Gamma_\phi^{(2)} M_{\rm Pl}} \simeq  8.8\times 10^{7} 
\langle \partial_\phi Y_{ijk} \rangle
\left(\frac{m_\phi}{10^{13}\, {\rm GeV}}\right)^{3/2} 
\,{\rm GeV},
\end{align}
with $g_\ast =106.75$, from which the reheating temperature is 
determined by the first derivative of the Yukawa coupling with respect 
to the inflaton field.


As a result, 
the relevant inflaton couplings are different from those of K\"ahler moduli 
and correspondingly one can distinguish the thermal history of the Universe after the inflation 
from that given by the K\"ahler moduli inflation. 
We will study these phenomenological aspects through  concrete models 
elsewhere.


\section{Conclusion}
\label{sec:con}
In this paper, we have studied the axion inflation with complex structure 
moduli within the framework of type IIB string theory on the Calabi-Yau manifold. 
In the usual moduli stabilization procedure in type IIB string theory, 
all of the complex structure moduli and dilaton are stabilized at the tree level in terms of 
three-form fluxes, whereas the K\"ahler moduli are stabilized at the nonperturbative 
level. 
However, along the line of Ref.~\cite{Hebecker:2015rya}, here we also have pointed out that 
the relevant complex structure 
moduli can be stabilized at the minimum through the instatonic correction in the 
period vector of mirror Calabi-Yau manifold. Then, the masses of these complex structure moduli 
could be lower than those of K\"ahler moduli. 
Based on this setup, we find the new type of axion inflation 
associated with the lightest complex structure modulus and its scalar potential is the mixture of 
polynomial functions and sinusoidal functions.
The new type of axion inflation is favored by the recent Planck data. 

The complex structure moduli inflation is different from K\"ahler moduli 
inflation in the light of the thermal history of the Universe after the inflation. 
Since the couplings between the complex structure moduli and 
the matter fields in the visible sector are different from those of K\"ahler moduli, 
the inflaton decay channel and the related reheating process are completely changed. 
Although the reheating process is highly model dependent, we show the typical decay 
channel and reheating temperature. 
It would be an important direction to construct the realistic phenomenological model 
combined with the complex structure moduli inflation. That will be studied elsewhere.

\subsection*{Acknowledgements}
The authors thank H.~Abe for useful discussions. 
T.~K. was
supported in part by
the Grant-in-Aid for Scientific Research No.~25400252 and No.~26247042 from the MEXT in
Japan.
H.~O. was supported in part by a Grant-in-Aid for JSPS Fellows 
No. 26-7296.

\end{document}